\documentstyle[12pt]{article}
\newcommand{\h}{\hspace*{.5mm}}

\newcommand{\hh}{\hspace*{10mm}}
\newcommand{\hhh}{\hspace*{15mm}}

\newcommand{\ba}{\begin{array}}
\newcommand{\ea}{\end{array}}
\newcommand{\bea}{\begin{eqnarray}}
\newcommand{\eea}{\end{eqnarray}}
\newcommand{\nn}{\nonumber}

\begin{document}

\begin{titlepage}
\newpage
\setcounter{page}{0}
\null
\rightline{DTP-TMI-007/95}

\rightline{October 1995}

\vspace{2cm}

\begin{center}
{\Large {\bf On the Meissner Effect in the Relativistic
\vspace{0.5cm}

Anyon Superconductors}}

\vspace{2cm}

{\large M.Eliashvili and G.Tsitsishvili}
\vspace{0.5cm}

{\it Department of Theoretical Physics \\
Tbilisi Mathematical Institute\\
Tbilisi 380093 Georgia\\
E-mail: simi@imath.kheta.ge}

\end{center}

\vspace{2cm}

\centerline{\bf Abstract}

\indent

The relativistic model with two types of planar fermions interacting with the
Chern-Simons and Maxwell fields is applied
to the study of anyon superconductor. It is demonstrated, that the Meissner
effect can be realized in the case of the simultaneous presence of the fermions
with a different magnetic moment interactions. Under the certain conditions
there occures an extra plateau at the magnetization curve. In the order under
consideration the spectrum of the electromagnetic field excitations contains
the long-range interaction and one massive "photon" state.

\end{titlepage}

\parindent 0mm
{\bf1. Introduction}
\parindent 8mm
\vspace*{5mm}

The zero-temperature Meissner effect presented in the 2+1 dimensional
anyon matter, provoked a considerable efforts in order to
promote the Chern-Simons gauge theory as a hypothetical
candidate for the high $T_c$ superconductivity.

The most important points in that development are existence
of the massless (Goldstone) pole in the current correlators
\cite {Fetter}, cancelation of bare and induced C-S terms \cite{Banks},
and detailed calculations of effective action and thermodynamical
potential for the fermions ineracting with C-S and Maxwell fields
\cite{Chen, Daemi, Hetrick, Lykken}.

Among the others it was shown, that the Meissner effect is partial, i.e.
magnetic field starts to penetrate into the sample at any non-zero
temperature \cite{Hetrick}. The second result concernes the effective mixing
of Maxwell and C-S fields and occurence of two distinct massive
excitations \cite{Daemi}.

In the present paper we try to give some complementary insights
into those intriguing questions.

It is demonstrated, that the Meissner effect exists only if the matter
consists of the two types of fermions with opposite signs of magnetic moment
interaction. Such a system can be naturally realized considering
the planar relativistic fermions. Note, that the different versions of the
relativistic anyon superconductivity have been considered in
\cite{Lykken, Dorey, Tsitsishvili}.

Studying the  gauge field propagators, we found out, that results
cited in \cite{Daemi}, can be understood as a consequenses of
momentum expansion of the certain structure functions.
Rising the order of  terms  taken into account, it can be shown,
that there exists the single massive excitation.

The paper is organized in the following way. Section 2 containes
the description of model and some basic definitions.

In Section 3 we present the thermodynamical potential and relevant
physical quantities.

In Section 4 we show, that the Meissner effect can take place if the system
contains two types of fermions. Further we calculate
the dependence of critical magnetic field on the temperature
and establish the existence of additional plateaux at the
magnetization curves.

In Section 5 we derive the electromagnetic field propagators and analyze
the spectrum.

\vspace*{15mm}
\parindent 0mm
{\bf2. The Model}
\parindent 8mm
\vspace*{5mm}

Consider the system of Dirac fermions in 2+1 dimensions interacting with
Maxwell electromagnetic field and Chern-Simons gauge field.

The total lagrangian for this system is a sum of matter and gauge field
lagrangians
\bea
{\cal L}=-\h\frac{\delta}{4\mu_0}\h F_{\mu\nu}F^{\mu\nu}
-\h\frac{c}{\hbar}\h\frac{e^2\nu_0}{2\pi}
\h\varepsilon^{\mu\nu\lambda}a_{\mu}\partial_{\nu}a_{\lambda}+ecn_eA_0+\nn
\eea
\bea
+\bar\psi\{i\h\hbar c\gamma^{\mu}D_{\mu}-\sigma mc^2\}\psi,\nn
\eea
\bea
D_{\mu}=\partial_{\mu}+i(e/\hbar)(A_{\mu}+a_{\mu}),\hhh m>0,\hhh\sigma=\pm1,\nn
\eea
where $\delta\sim10^{-9}m$ is the interplanar distance in the multilayered
system and its presence is justified by the use of the three dimensional
Maxwell field and electric charge.
The Chern-Simons field is in fact two-dimensional, but will be
measured in three dimensional units.
$\psi$ is the 2-component fermion field and $en_e$ is the planar density
of the background neutralizing charges. In 2+1 dimensions the sign of the mass
term is essential and describes the helicity of the fermion.

In what follows, we study the static properties of this relativistic system.
Analysis will be performed in the self consistent field approximation,
developed in \cite{Hetrick}. It means, that the gauge fields must satisfy the
equations
\bea
-\h\frac{\delta}{\mu_0}\h\Delta A_0+ecn_e=\langle J^0({\bf r})\rangle,
\eea
\bea
-\h\frac{\delta}{\mu_0}\h\varepsilon^{kn}\partial_nB
=\langle J^k({\bf r})\rangle,
\eea
\bea
\frac{c}{\hbar}\h\h\frac{e^2\nu_0}{\pi}\h\h b=\langle J^0({\bf r})\rangle,
\eea
\bea
\frac{c}{\hbar}\h\h\frac{e^2\nu_0}{\pi}\h\h\varepsilon^{kn}\partial_n a_0
=\langle J^k({\bf r})\rangle,
\eea
where the thermal averages are defined in terms of the grand canonical ensamble
\bea
\langle\cdots\rangle\h{\rm Tr}e^{-(H_e-\mu mc^2N)/k_BT}
={\rm Tr}\left\{\cdots e^{-(H_e-\mu mc^2N)/k_BT}\right\}.\nn
\eea
Here $k_B$, $T$ and $\mu$ are the Boltzmann constant, temperature and the
dimensionless chemical potential respectively, while $N$ denotes the particle
number operator
\bea
N=\int\psi^{\dagger}({\bf r})\psi({\bf r}){\rm d}{\bf r}.\nn
\eea
The fermion hamiltonian is given by
\bea
H_e=\int\psi^{\dagger}({\bf r})\{i\hbar c\gamma^0\gamma_kD_k({\bf r})
+ecA_0({\bf r})+eca_0({\bf r})
+\sigma mc^2\gamma^0\}\psi({\bf r}){\rm d}{\bf r}.\nn
\eea

Introduce the thermodynamic potential
\bea
\Omega_e(T,\h\mu,\h A,\h a)=-k_BT\ln{\rm Tr}\exp
\left\{-\h\frac{H_e(A,\h a)-\mu mc^2N}{k_BT}\right\},\nn
\eea
in terms of which the current averages are given by
\bea
\langle\h J^{\mu}({\bf r})\h\rangle
=\frac{\delta\Omega_e}{\delta A_{\mu}({\bf r})}
=\frac{\delta\Omega_e}{\delta a_{\mu}({\bf r})}.
\eea

The gauge fields are presented as a sums of background and fluctuating parts
\bea
A^{}_{\mu}=A^{\rm b}_{\mu}+A^{\rm f}_{\mu},\hh\hh
a^{}_{\mu}=a^{\rm b}_{\mu}+a^{\rm f}_{\mu}.\nn
\eea
Backgrounds correspond to the uniform Maxwell and Chern-Simons magnetic fields,
i.e. $A^{\rm b}_0=a^{\rm b}_0=0$, $B^{\rm b}=const$, $b^{\rm b}=const$.

Separate the fermionic hamiltonian into the dimensionless free and interacting
parts
\bea
H_e=mc^2(H_0+H_{int}),\nn
\eea
\bea H_0
=\int\psi^{\dagger}({\bf r})\{i\h\ell_0\gamma^0\gamma^{}_kD^{\rm b}_k({\bf r})
+\sigma\gamma^0\}\psi({\bf r}){\rm d}{\bf r},
\eea
\bea
D^{\rm b}_k({\bf r})
=\partial^{}_k+i(e/\hbar)(A^{\rm b}_k+a^{\rm b}_k),\nn
\eea
\bea
H_{int}=\frac{1}{mc^2}\int J^{\mu}({\bf r})
\{A^{\rm f}_{\mu}({\bf r})+a^{\rm f}_{\mu}({\bf r})\}{\rm d}{\bf r},\nn
\eea
where $\ell_0=\hbar/mc$ is the Compton wave length for the fermions.

Applying the perturbation theory formalism, we get for the thermodynamic
potential
\bea
\Omega_e=\Omega_0-k_BT\ln\left\langle{\rm T}\exp
\left\{-\int_0^{\beta}H_{int}(\tau)d\tau\right\}\right\rangle_0,
\hhh\beta\h=\h\h\frac{mc^2}{k_BT}.
\eea
$\Omega_0$ is the thermodynamic potential for the system in the uniform
magnetic background
\bea
\Omega_0(T,\h\mu,\h A^{\rm b},\h a^{\rm b})=-k_BT\ln{\rm Tr}\exp
\left\{-\h\beta(H_0(A^{\rm b},\h a^{\rm b})-\mu N)\right\}
\eea
and $\langle\cdots\rangle^{}_0$ is defined as
\bea
\langle\cdots\rangle^{}_0{\rm Tr}\left\{e^{-\beta(H_0-\mu N)}\right\}
={\rm Tr}\left\{\cdots e^{-\beta(H_0-\mu N)}\right\}.\nn
\eea

In (7) $H_{int}(\tau)$ is the interaction hamiltonian in the Matsubara
representation
\bea
H_{int}(\tau)=\frac{1}{mc^2}\int\bar\psi(\tau,\h{\bf r})\gamma^{\mu}
\psi(\tau,\h{\bf r})
\{A^{\rm f}_{\mu}({\bf r})+a^{\rm f}_{\mu}({\bf r})\}{\rm d}{\bf r}\nn
\eea
and T denotes $\tau$-ordering. Matsubara fields are given by
\bea
\psi(\tau,\h{\bf r})=e^{\mu\tau}\sum_{np}
\{a^{}_{np}u^{}_{np}({\bf r})e^{-\omega_n\tau}+
b^{\dagger}_{np}v^{}_{np}({\bf r})e^{\omega_n\tau}\},
\eea
\bea
\bar\psi(\tau,\h{\bf r})=e^{-\mu\tau}\sum_{np}
\{a^{\dagger}_{np}\bar u^{}_{np}({\bf r})e^{\omega_n\tau}+
b^{}_{np}\bar v^{}_{np}({\bf r})e^{-\omega_n\tau}\}.
\eea

Remark, that except the lowest energy eigenvalues the solutions of
Dirac equation are always paired. For the lowest energy there is asymmetry,
i.e. there is no $v_0$ mode for $\sigma\varepsilon=+1$ and
no $u_0$ mode for $\sigma\varepsilon=-1$, where
$\varepsilon\equiv{\rm sgn}(eB^{\rm b}+eb^{\rm b})$.

\vspace*{15mm}
\parindent 0mm
{\bf3. Thermodynamic potential}
\parindent 8mm
\vspace*{5mm}

In this item we shall find the analytical expression for the thermodynamic
potential in the second order approximation with respect to the gauge field
fluctuations.  All the operator expressions such as hamiltonian, currents,
{\it etc.} are assumed to be normal ordered.

Substituting $H_0$ into (8) and taking into account the degeneracy of the
Landau levels we get
\bea
\frac{\Omega_0(A^{\rm b},\h a^{\rm b})}{Area}
=\frac{k_BT}{2\pi\ell^2}\sum_{n}{\rm ln}(1-\rho^{+}_n)
+\frac{k_BT}{2\pi\ell^2}\sum_{n}{\rm ln}(1-\rho^{-}_n).
\eea
where $\rho^+_n$ and $\rho^-_n$ are the Fermi distribution functions for the
particles and antiparticles respectively
\bea
\rho^{\pm}_n=\left\{1+\exp[\h\beta(\omega_n\mp\mu)]\right\}^{-1},
\hhh\omega_n=\sqrt{1+2hn},\nn
\eea
\bea h\h=\h\h\frac{\ell_0^2}{\ell^2},\hhh\hhh
\frac{1}{\ell^2}\h\h=\h\h\frac{1}{\hbar}\h\h|eB^{\rm b}\h+\h eb^{\rm b}|
\eea
and $\ell$ is the magnetic length.

Thermodynamic potential in the second order approximation is given by the
functional
\bea
\Omega_e(A,\h a)=\Omega_0(A^{\rm b},\h a^{\rm b})+ec\int\Pi^{\mu}({\bf r})
\{A^{\rm f}_{\mu}({\bf r})+a^{\rm f}_{\mu}({\bf r})\}{\rm d}{\bf r}+\nn
\eea
\bea
+\frac{e^2}{2m}\int\Pi^{\mu\nu}({\bf r}_1,\h{\bf r}_2)
\{A^{\rm f}_{\mu}({\bf r}_1)+a^{\rm f}_{\mu}({\bf r}_1)\}
\{A^{\rm f}_{\nu}({\bf r}_2)+a^{\rm f}_{\nu}({\bf r}_2)\}
{\rm d}{\bf r}_1{\rm d}{\bf r}_2,\nn
\eea
where $\Pi^{\mu}({\bf r})$ and $\Pi^{\mu\nu}({\bf r}_1,\h{\bf r}_2)$ read as
\bea
\Pi^{\mu}({\bf r})
=\frac{1}{2\pi\ell^2}\sum_n\rho^+_n\bar u^{}_n\gamma^{\mu}u^{}_n
-\frac{1}{2\pi\ell^2}\sum_n\rho^-_n\bar v^{}_n\gamma^{\mu}v^{}_n=const,\nn
\eea
\bea
\Pi^{\mu\nu}({\bf r}_1,\h{\bf r}_2)=
\frac{1}{\ell^4}\int\frac{{\rm d}{\bf k}}{8\pi^3}\h
e^{i{\bf k}({\bf r}_1-{\bf r}_2)/\ell}\Pi^{\mu\nu}({\bf k}).\nn
\eea

Polarization operator possesses the following tensor form
\bea
\Pi^{00}({\bf k})=
\h\frac{2\pi}{h}\h\Pi_{\rm E}({\bf k}^2/2),\nn
\eea
\bea
\Pi^{0k}({\bf k})=
\h\frac{2\pi}{\sqrt h}\h i\varepsilon^{kl}k^l\Pi_{\rm CS}({\bf k}^2/2),\nn
\eea
\bea
\Pi^{kl}({\bf k})=
2\pi\varepsilon^{km}\varepsilon^{ln}k^mk^n\Pi_{\rm M}({\bf k}^2/2),\nn
\eea
where $\Pi_{\rm E}$, $\Pi_{\rm CS}$ and $\Pi_{\rm M}$ are the structure
functions, presented in the Appendix.

Explicit calculations show that
\bea
\Pi_0
=\h\h\frac{1}{2\pi\ell^2}\sum_n\rho^{+}_n
-\h\h\frac{1}{2\pi\ell^2}\sum_n\rho^{-}_n,
\hh\hh\Pi_k=0
\eea
and the final expression for the thermodynamic potential is given by
\bea
\Omega_e(A,\h a)=\Omega^{}_0(A^{\rm b},\h a^{\rm b})+ec\h\Pi^{}_0
\int\{A^{\rm f}_0({\bf r})+a^{\rm f}_0({\bf r})\}{\rm d}{\bf r}+\nn
\eea
\bea
+\frac{e^2}{2m}\h\frac{m^2c^2}{\hbar^2}\int
\{A^{\rm f}_0({\bf r})+a^{\rm f}_0({\bf r})\}\hat\Pi^{}_{\rm E}
\{A^{\rm f}_0({\bf r})+a^{\rm f}_0({\bf r})\}{\rm d}{\bf r}+\nn
\eea
\bea
+\frac{e^2}{2m}\h\frac{2mc}{\hbar^2}\int
\{A^{\rm f}_0({\bf r})+a^{\rm f}_0({\bf r})\}\hat\Pi^{}_{\rm CS}
\{B^{\rm f}({\bf r})+b^{\rm f}({\bf r})\}{\rm d}{\bf r}+\nn
\eea
\bea
+\frac{e^2}{2m}\int
\{B^{\rm f}({\bf r})+b^{\rm f}({\bf r})\}\hat\Pi^{}_{\rm M}
\{B^{\rm f}({\bf r})+b^{\rm f}({\bf r})\}{\rm d}{\bf r},\nn
\eea
where $\hat\Pi$ stands for the differential operator $\Pi(-\ell^2\Delta/2)$
($\Delta$ is the Laplace operator). Using (5), the current averages can
be expressed as a linear functions of gauge field fluctuations
\bea
\langle\h J^0({\bf r})\h\rangle
=ec\h\Pi_0
+\h\h\frac{e^2mc^2}{\hbar^2}\h\h\hat\Pi_{\rm E}
\left(A^{\rm f}_0+a^{\rm f}_0\h\right)
+\h\h\frac{e^2c}{\hbar}\h\h\hat\Pi_{\rm CS}\left(B^{\rm f}+b^{\rm f}\h\right),
\eea
\bea
\langle\h J^k({\bf r})\h\rangle
=-\h\varepsilon^{kn}\h\partial_n\left\{
\frac{e^2c}{\hbar}\h\h\hat\Pi_{\rm CS}\left(A^{\rm f}_0+a^{\rm f}_0\h\right)
+\frac{e^2}{m}\h\h\hat\Pi_{\rm M}\left(B^{\rm f}+b^{\rm f}\h\right)\right\}.
\eea

\vspace*{15mm}
\parindent 0mm
{\bf 4. Uniform magnetic field and the Meissner effect}
\parindent 8mm
\vspace*{5mm}

As a starting point, consider the system in the zeroth order approximation,
i.e. $A^{\rm f}_{\mu}=a^{\rm f}_{\mu}=0$. Equations of motion (1) and (3) are
reduced to
\bea
n_e=\Pi_0,
\eea
\bea
b^{\rm b}=\h\h\frac{\pi\hbar}{e\nu_0}\h\h n_e,
\eea
where $\Pi_0$ is given by (13).

Note, that $n_e$ is the free fermion density in the sample and (16) serves to
define the chemical potential $\mu=\mu(T,B^{\rm b},b^{\rm b},n_e)$.

Equation (17) indicates that the Chern-Simons magnetic background is created by
the free fermion density. For the typical value $n_e=10^{18}{\rm m}^{-2}$ we
get
\bea
|\h b^{\rm b}|=\h\h\frac{2\cdot10^7}{|\h\nu_0|}\h\h Gauss.
\eea

We shall study the compound system, consisting of two sorts of fermions with
equal gauge couplings $e_1=e_2=e$ and the different helicities
$\sigma_1=-\h\sigma_2$, corresponding to the different signs of the magnetic
moment interaction, which on its turn can be associated with spin up and spin
down fermions. In that case the r.h.s. of the equation (16) gets the
contributions, defined by (13) from both sorts of particles.
Remark that, since the different sorts have the equal charges, corresponding
magnetic lengths are also equal. Moreover, as it follows from (18), realistic
values of $B^{\rm b}$ are small compared with those of $b^{\rm b}$ and can be
neglected:
\bea
\varepsilon={\rm sgn}(eB^{\rm b}+eb^{\rm b})={\rm sgn}(eb^{\rm b}).\nn
\eea
In other words, $\varepsilon$ can be considered as independent of $B^{\rm b}$,
and without loss of generality we can set
$\sigma_1\varepsilon_1=-\sigma_2\varepsilon_2=1$. Taking into account the
contributions from both types, equation of motion (16) can be rewritten in the
following form
\bea
n_e=\h\h\frac{h}{2\pi\ell_0^2}\h\h(\nu_1+\nu_2),
\eea
where $\nu_1$ and $\nu_2$ are the corresponding filling fractions
\bea
\nu_1\equiv\sum_{n=0}\rho^+_n(\mu_1)-\sum_{n=1}\rho^-_n(\mu_1),
\eea
\bea
\nu_2\equiv\sum_{n=1}\rho^+_n(\mu_2)-\sum_{n=0}\rho^-_n(\mu_2),
\eea
and $\mu_{1,2}$ are the chemical potentials for the types 1 and 2 respectively.
In (20) and (21) we took into the consideration the spectral asymmetry of the
one-particle hamiltonian, which is reflected in the absence of $n=0$ modes in
certain terms.

Introduce the partial contributions to the particle density
\bea
n^{(1,2)}_e\h=\h\h\frac{h}{2\pi\ell_0^2}\h\h\nu_{1,2}.
\eea

By means of (19) we can express $h$ in terms of the average filling fraction
\bea
h\h=\h\h\frac{\pi n_e\ell_0^2}{\nu},\hhh\hh
\nu\h=\h\h\frac{\nu_1\h+\nu_2}{2}.
\eea
Using (22), (23), we express the partial filling fractions in terms of the
average one
\bea
\nu_{1,2}\h=\h\h\frac{2\h n_e^{(1,2)}}{n_e^{(1)}\h+\h n_e^{(2)}}\h\h\nu.
\eea
Substituting $h$, $\nu_{1,2}$ from (23) and (22) into (20) and (21) we see
that the chemical potentials depend on temperature and the average filling
fraction $\nu$.

The same time, expression for the magnetic length (12) together with equations
(17) and (23) yields
\bea
\frac{1}{\nu}\h\h=\h\h\frac{\varepsilon}{\nu_0}\h\h
+\h\h\frac{\varepsilon e B^{\rm b}}{\pi n_e\hbar},
\eea
which reflects the one to one correspondence between $\nu$ and $B^{\rm b}$.
Consequently, any quantity depending on $B^{\rm b}$ can be also viewed as a
function of $\nu$, and {\it vice versa}.

The value of the background magnetic field is determined by the external
magnetic field $B^{\rm ext}$ and the magnetization $M$
\bea
B^{\rm b}=B^{\rm ext}+M(B^{\rm b}),
\hh\hh M(B^{\rm b})
=-\h\h\frac{\mu_0}{\delta}\h\h\frac{d{\cal F}(B^{\rm b})}{dB^{\rm b}},
\eea
where ${\cal F}\h=\h{\cal F}_1\h+\h{\cal F}_2$ is the Helmholtz free energy
density of the composite system. The separate contributions are
\bea
{\cal F}_{1,2}(B^{\rm b})\h=\h\h\frac{1}{Area}
\h\h\{\h\Omega_0(\mu_{1,2})
\h+\h\mu_{1,2}\h mc^2\langle\h N_{1,2}\h\rangle\},\nn
\eea
where $\langle\h N_{1,2}\h\rangle$ are the thermal averages of the fermion
numbers determined by
\bea
\frac{\langle\h N_1\h\rangle}{Area}\h\h
=\h\h\frac{h}{2\pi\ell^2_0}\sum_{n=0}\rho^+_n(\mu_1)
-\h\h\frac{h}{2\pi\ell^2_0}\sum_{n=1}\rho^-_n(\mu_1)
=\h n_e^{(1)},\nn
\eea
\bea
\frac{\langle\h N_2\h\rangle}{Area}\h\h
=\h\h\frac{h}{2\pi\ell^2_0}\sum_{n=1}\rho^+_n(\mu_2)
-\h\h\frac{h}{2\pi\ell^2_0}\sum_{n=0}\rho^-_n(\mu_2)
=\h n_e^{(2)}.\nn
\eea

Here the quantities $h$, $\mu_1$ and $\mu_2$ are the functions of $\nu$.
In order to exhibit the global behaviour of ${\cal F}({\nu})$ we have used the
numerical methods. Consider the case of the equal concentrations
$n^{(1)}_e=n^{(2)}_e=n_e/2$. Corresponding results are plotted in figure 1. As
one can see the free energy density of the composite system has a local minima
at the integer values of the average filling fraction, however the separate
contributions of each type of particles exhibit no minima.

The cusp like structure of the free energy is the manifestation of the Meissner
effect, when the system tryies to expell the magnetic field from inside the
sample \cite{Hetrick}.

As we see, in contrast with the previous calculations, the Meissner effect does
not exist in the single fermion system, but only in the composite one, where
the diversity in the magnetic moment interaction plays the decisive role.

Magnetization is expressed with the help of the free energy. The later
containes the additive contributions from the one-particle state energies.
One particle hamiltonian considered in \cite{Hetrick}, takes into account the
magnetic moment interaction only with Maxwell magnetic field.
Further simplification is achieved by taking
\bea
E_n=\frac{\hbar^2}{m\ell^2}\left(n+\frac{1}{2}\right),
\eea
assuming the vanishing magnetic moment interaction.

In our approximation the magnetic interaction term containes the contribution
from both Maxwell and Chern-Simons magnetic fields. The nonrelativistic limit
of one-particle energy spectrum is given by
\bea
E_n(\sigma\varepsilon)
=\frac{\hbar^2}{m\ell^2}\left(n+\frac{1-\sigma\varepsilon}{2}\right).\nn
\eea
Remark, that (27) is in fact the half sum of $E_n(+)$ and $E_n(-)$. Separate
use of $E_n(+)$ or $E_n(-)$ does not lead to the free energy with localized
minima and only their simultaneous contribution has a cusp like form.

Equations for the chemical potentials cannot be solved in the global form.
However, one can find the analytic form of $\mu_{1,2}(\nu)$ nearby the integer
values of $\nu$, where the free energy achieves the local minima. This enables
to analyze the system in more details near these minima, where the Meissner
effect just takes the place. Below we present these calculations for the
asymmetric concentrations i.e. when $n^{(1)}_e\neq n^{(2)}_e$.

Here we shall deal with the minimum corresponding to $\nu=1$ and for
concretness set $e>0$ and $\varepsilon=1$. In that case we have $\nu_0=1$ and
\bea
h\h=\h\pi n_e\ell_0^2(1\h+\h\alpha),\hh\hh
\alpha\h=\h\h\frac{eB}{\pi n_e\hbar}
\eea
Note that for the characteristic values of the internal magnetic field
$(B<200\h Gauss)$ we have $\alpha<10^{-5}$, $(1+\alpha)^{-1}=1-\alpha$ and
present (24) as
\bea
\nu_1\h=\h(1\h+\h\bar\alpha)(1\h-\h\alpha)\h\equiv\h1\h-\h\alpha_1,\nn
\eea
\bea
\nu_2\h=\h(1\h-\h\bar\alpha)(1\h-\h\alpha)\h\equiv\h1\h-\h\alpha_2,\nn
\eea
\bea
\bar\alpha\h=\h\h\frac{n^{(1)}_e-\h\h n^{(2)}_e}{n^{(1)}_e+\h\h n^{(2)}_e}.\nn
\eea
Here $\bar\alpha$ measures the asymmerty in the concentrations of the different
types of fermions.

Further simplifications are due to the fact that in the considered range of
temperatures we have $\beta>3\cdot10^7$, and consequently for $\mu_{1,2}>0$
\bea
\rho^{-}_{n}(\mu_{1,2})
=\frac{1}{1+e^{\beta(\omega_n+\mu_{1,2})}}<e^{-10^7},\nn
\eea
meaning that the main contributions to (20) and (21) come from
$\rho^+_n(\mu_{1,2})$, forcing $\nu_{1,2}$ to be positive.
With this assumption we take $\mu_1>0$ and present it as
\bea
\mu_1\h=\h1\h+\h\h\frac{1\h-\h w_1}{2}\h\h h,\nn
\eea
where $w_1$ is to be found. Due to (28) one has $h\sim5\cdot10^{-7}$.
Consequently, for the Landau levels with $2hn<<1$ we can use $\omega_n=1+hn$
and write down
\bea
\beta(\omega_n\h-\mu_1)\h
=\h\beta h\left(n\h-\h\h\frac12\h\h+\h\h\frac{w_1}{2}\right).\nn
\eea

Characteristic values of $\beta$ and $h$ are such that $\beta h>15$, allowing
to neglect the contributions coming from the higher Landau levels.
This permits to write (20) as $\nu_1=\rho^{+}_{0}(\mu_1)+\rho^{+}_{1}(\mu_1)$
or in the equivalent form
\bea
1\h-\h\alpha_1\h
=\h\h\frac{1}{1+e^{-\eta}e^{\eta w_1}}
\h\h+\h\h\frac{1}{1+e^{\eta}e^{\eta w_1}},\nn
\eea
where $\eta\equiv\pi n_e\ell_0^2\beta(1+\alpha)/2$. From this equation we
easily solve $w_1$ as
\bea
e^{\pm\eta w_1}\h=\h\h\frac{1}{1\mp\alpha_1}
\left(\sqrt{1\h+\h\alpha_1^2{\rm sh}^2\eta}\h\pm\h\alpha_1{\rm ch}\eta\right),
\hhh\alpha_1\h=\h\alpha\h-\h\bar\alpha\h+\h\alpha\bar\alpha.
\eea

The same time, we take
\bea
\mu_2\h=\h1+\h\h\frac{3-w_2}{2}\h\h h\nn
\eea
and performing the same manipulations obtain
\bea
e^{\pm\eta w_2}\h=\h\h\frac{1}{1\mp\alpha_2}
\left(\sqrt{1\h+\h\alpha_2^2{\rm sh}^2\eta}\h\pm\h\alpha_2{\rm ch}\eta\right),
\hhh\alpha_2\h=\h\alpha\h+\h\bar\alpha\h-\h\alpha\bar\alpha.
\eea

So, the leading contributions to the different physical quantities come from
the following Fermi distribution functions
\bea
\rho^+_0(\mu_1)\h=\h\h\frac{1}{1+e^{-\eta}e^{\eta w_1}},\hh\hh
\rho^+_1(\mu_1)\h=\h\h\frac{1}{1+e^{\eta}e^{\eta w_1}},
\eea
\bea
\rho^+_1(\mu_2)\h=\h\h\frac{1}{1+e^{-\eta}e^{\eta w_2}},\hh\hh
\rho^+_2(\mu_2)\h=\h\h\frac{1}{1+e^{\eta}e^{\eta w_2}},
\eea
where $e^{\eta w_1}$ and $e^{\eta w_2}$ are defined by (29) and (30).

In this approximation the system magnetization is $M\h=\h M_1\h+\h M_2$ where
\bea
M_1(B^{\rm b})\h=\h-\h\h\frac{e\mu_0mc^2}{2\pi\hbar\delta}
\sum_{n=0,1}\left\{\frac{1}{\beta}\h\h{\rm ln}\h[\h1-\rho^{+}_{n}(\mu_1)]
+\h\h\frac{hn}{\omega_n}\h\h\rho^{+}_{n}(\mu_1)\right\}\approx\nn
\eea
\bea
\approx\h\h\frac{\mu_0en_e\hbar}{4m\delta}
\left\{f(\alpha\h-\h\bar\alpha)\h+\h1\right\},\hspace*{31mm}
\eea
\bea
M_2(B^{\rm b})\h=\h-\h\h\frac{e\mu_0mc^2}{2\pi\hbar\delta}
\sum_{n=1,2}\left\{\frac{1}{\beta}\h\h{\rm ln}\h[\h1-\rho^{+}_{n}(\mu_2)]
+\h\h\frac{hn}{\omega_n}\h\h\rho^{+}_{n}(\mu_2)\right\}\approx\nn
\eea
\bea\approx\h\h\frac{\mu_0en_e\hbar}{4m\delta}
\left\{f(\alpha\h+\h\bar\alpha)\h-\h1\right\},\hspace*{31mm}
\eea
\bea
f(z)\h=\h\h\frac{1}{\xi}\h\h{\rm ln}\h\h\frac
{\sqrt{z^2\h+\h4\h e^{-\xi}}\h-\h z}
{\sqrt{z^2\h+\h4\h e^{-\xi}}\h+\h z},\hh\hh\xi=\pi n_e\ell_0^2\beta
\eea
These quantities define the magnetic and thermal properties of the system.

In the case of the symmetric concentrations $(\bar\alpha=0)$ one gets
\bea
M(T,B^{\rm b})\h
=\h\h\frac{\mu_0en_e\hbar}{2m\delta}\h\h\frac{1}{\xi}\h\h{\rm ln}
\h\h\frac{\sqrt{\alpha^2\h+\h4\h e^{-\xi}}\h-\h\alpha}
         {\sqrt{\alpha^2\h+\h4\h e^{-\xi}}\h+\h\alpha},
\hh\hh\alpha\h=\h\h\frac{eB^{\rm b}}{\pi n_e\hbar}.\nn
\eea
Behaviour of $M$ as a function of $B^{\rm b}$ is depicted in figure 2 for the
different values of temperature. Presented curves are the same as given in
earlier work \cite{Hetrick}.

On figure 3 we present the dependence of $B^{\rm b}$ on $B^{\rm ext}$ which is
set by (26). As one can see, when the external magnetic field is applyied, the
internal one is in general non-zero. In other words, the Meissner effect is not
complete. This observation was first made in \cite{Hetrick}. The same time the
magnitude of $B^{\rm b}$ is quite small until $B^{\rm ext}$ reaches some
critical value $B^{\rm cr}(T)$, which depends on the temperature. Above this
value external magnetic field begins the notable penetration in the sample.

The critical value of the external magnetic field is evidently related to the
small interval at $B^{\rm b}$ axis, where the magnetization curve drastically
changes its direction, i.e. where the curve passes the point of maximal
curvature (PMC). The lower is temperature, more narrow is the interval and it
is easier to establish the corresponding critical magnetic field. In order to
find the approximated value of $B^{\rm cr}(T)$ we can use the behaviour of the
derivative
\bea
\frac{\partial M}{\partial B^{\rm b}}
=-\h\h\frac{\mu_0e^2k_BT}{\pi^2\hbar^2n_e\delta}
\h\h\left(\alpha^2+4e^{-\pi n_e\hbar^2/mk_BT}\right)^{-1/2}.\nn
\eea

Until the curve $M(B^{\rm b})$ reaches the PMC, its slope can be considered to
be constant and therefore can be determined by its value at the origin, where
it is of the order of $10^5$ for $T=50^{\circ}K$ and $10^{50}$ for
$T=10^{\circ}K$.  On the other hand, in the low temperature regime the curve
becomes practically horizontal after passing the PMC. Obviously, somewhere in
the vicinity of PMC one has $\partial M/\partial B^{\rm b}=-1$ and using this
relation as the definition of the location of PMC, one gets the corresponding
value of the internal magnetic field to be
\bea
B_0(T)
=\left\{\left(\frac{\mu_0ek_BT}{\pi\hbar\delta}\right)^2
-\left(\frac{2\pi n_e\hbar}{e}\right)^2 \h\h e^{-\pi
n_e\hbar^2/mk_BT}\right\}^{1/2}.\nn
\eea

Substituting $B_0(T)$ into $B^{\rm cr}=B_0-M(B_0)$ and keeping the leading
terms we get
\bea
B^{\rm cr}(T)=\h\h\frac{\mu_0en_e\hbar}{2m\delta}\h\h
+\h\h\frac{\mu_0ek_BT}{\pi\hbar\delta}\h\h{\rm ln}
\h\h\frac{\mu_0e^2k_BT}{2\pi^2\hbar^2n_e\delta}.\nn
\eea
Corresponding curve is presented at figure 4. We observe that the
value of the critical magnetic field decreases for higher temperatures
and completely vanishes near $T=90^{\circ}K$, which seems to be the critical
temperature for the system. However, the area of the maximal curvature becomes
smeared for $T=90^{\circ}K$, and the critical magnetic field is ill defined,
causing the critical temperature to be also ill defined. On the contrary, the
PMC is well located for $T=0^{\circ}K$ and the corresponding value of the
critical magnetic field is given by
\bea
B^{\rm cr}(0)=\h\h\frac{\mu_0en_e\hbar}{2m\delta}.\nn
\eea

Situation with the asymmetric concentrations is presented at figures 5 and 6.
In the interval, which includes $B^{\rm b}=0$, magnetization practilally
vanishes. The width of the plateaux, is equal to
$2\bar\alpha\cdot(\pi n_e\hbar/e)$ and does not depend on temperature.
This kind of behaviour of the magnetization can be explained by the
fact, that the magnetization curves of each kind of fermions have the standard
form set by (35), but shifted by $\bar\alpha$ relatively to each other
according to (33) and (34).

The genuine value of the asymmetry parameter $\bar\alpha$ must be determined
by the equilibrium conditions imposed on the chemical potentials.

\vspace*{15mm}
\parindent 0mm
{\bf5. Gauge field propagators and mass spectrum}
\parindent 8mm
\vspace*{5mm}

Equations of motion for the gauge fluctuations can be introduced as the
extremals of the effective lagrangian
\bea
{\cal L}_{\rm eff}
\h=\h\h\frac{\delta}{2\mu_0}\h\h(\partial_kA^{\rm f}_0)^2
\h-\h\h\frac{e^2mc^2}{2\hbar^2}\h\h(A^{\rm f}_0+a^{\rm f}_0)
\h\hat\Pi^{\rm tot}_{\rm E}\h(A^{\rm f}_0+a^{\rm f}_0)\h-\nn
\eea
\bea
-\h\h\frac{\delta}{2\mu_0}\h\h B^{\rm f}B^{\rm f}
\h-\h\h\frac{e^2}{2m}\h\h(B^{\rm f}+b^{\rm f})
\h\hat\Pi^{\rm tot}_{\rm M}\h(B^{\rm f}+b^{\rm f})\h+\nn
\eea
\bea
+\h\h\frac{e^2\nu_0c}{\pi\hbar}\h\h a^{\rm f}_0b^{\rm f}
\h-\h\h\frac{e^2c}{\hbar}\h\h(A^{\rm f}_0+a^{\rm f}_0)
\h\hat\Pi^{\rm tot}_{\rm CS}\h(B^{\rm f}+b^{\rm f}),\nn
\eea
where the script "tot" means that the corresponding quantities include the
contributions from both types of fermions.

In order to study the gauge field propagators we introduce the gauge fixing
term
\bea
{\cal L}_{\alpha}
\h=\h\h\frac{1}{2\alpha}\h\h(\partial^{}_nA^{\rm f}_n)^2
\h+\h\h\frac{1}{2\alpha}\h\h(\partial^{}_na^{\rm f}_n)^2\nn
\eea
and present the total lagrangian as
\bea
{\cal L}_{\rm eff}\h+\h{\cal L}_{\alpha}\h
=\h\h\frac12\h\h{\cal A}^{\h\rm T}D\h{\cal A},
\hhh{\cal A}^{\h\rm T}=\left(\h ecA^{\rm f}_0\h,\h eA^{\rm f}_n\h,
\h eca^{\rm f}_0\h,\h ea^{\rm f}_n\h\right).\nn
\eea

Gauge field propagators are defined by the inverse of the matrix $D$
\bea
DG=GD=1\nn
\eea
and can be presented in the following form
\vglue .1cm
\bea
G\hspace*{5mm}=\hspace*{5mm}
\setlength{\unitlength}{.25mm}
\left\lgroup\h\h\h\h
\begin{picture}(264,152)

\multiput(0,136)(4,0){11}{\circle*{1}}
\multiput(0,96)(4,0){11}{\circle*{1}}
\multiput(0,96)(0,4){11}{\circle*{1}}
\multiput(40,96)(0,4){11}{\circle*{1}}
\put(8,110){$G^{00}_1$}

\multiput(48,136)(4,0){21}{\circle*{1}}
\multiput(48,96)(4,0){21}{\circle*{1}}
\multiput(48,96)(0,4){11}{\circle*{1}}
\multiput(128,96)(0,4){11}{\circle*{1}}
\put(74,110){$G^{0m}_1$}

\multiput(136,136)(4,0){11}{\circle*{1}}
\multiput(136,96)(4,0){11}{\circle*{1}}
\multiput(136,96)(0,4){11}{\circle*{1}}
\multiput(176,96)(0,4){11}{\circle*{1}}
\put(144,110){$G^{00}_3$}

\multiput(184,136)(4,0){21}{\circle*{1}}
\multiput(184,96)(4,0){21}{\circle*{1}}
\multiput(184,96)(0,4){11}{\circle*{1}}
\multiput(264,96)(0,4){11}{\circle*{1}}
\put(210,110){$G^{0m}_3$}

\multiput(0,88)(4,0){11}{\circle*{1}}
\multiput(0,8)(4,0){11}{\circle*{1}}
\multiput(0,8)(0,4){21}{\circle*{1}}
\multiput(40,8)(0,4){21}{\circle*{1}}
\put(8,42){$G^{n0}_1$}

\multiput(48,88)(4,0){21}{\circle*{1}}
\multiput(48,8)(4,0){21}{\circle*{1}}
\multiput(48,8)(0,4){21}{\circle*{1}}
\multiput(128,8)(0,4){21}{\circle*{1}}
\put(74,42){$G^{nm}_1$}

\multiput(136,88)(4,0){11}{\circle*{1}}
\multiput(136,8)(4,0){11}{\circle*{1}}
\multiput(136,8)(0,4){21}{\circle*{1}}
\multiput(176,8)(0,4){21}{\circle*{1}}
\put(144,42){$G^{n0}_4$}

\multiput(184,88)(4,0){21}{\circle*{1}}
\multiput(184,8)(4,0){21}{\circle*{1}}
\multiput(184,8)(0,4){21}{\circle*{1}}
\multiput(264,8)(0,4){21}{\circle*{1}}
\put(210,42){$G^{nm}_3$}

\multiput(0,0)(4,0){11}{\circle*{1}}
\multiput(0,-40)(4,0){11}{\circle*{1}}
\multiput(0,-40)(0,4){11}{\circle*{1}}
\multiput(40,-40)(0,4){11}{\circle*{1}}
\put(8,-26){$G^{00}_3$}

\multiput(48,0)(4,0){21}{\circle*{1}}
\multiput(48,-40)(4,0){21}{\circle*{1}}
\multiput(48,-40)(0,4){11}{\circle*{1}}
\multiput(128,-40)(0,4){11}{\circle*{1}}
\put(74,-26){$G^{0m}_4$}

\multiput(136,0)(4,0){11}{\circle*{1}}
\multiput(136,-40)(4,0){11}{\circle*{1}}
\multiput(136,-40)(0,4){11}{\circle*{1}}
\multiput(176,-40)(0,4){11}{\circle*{1}}
\put(144,-26){$G^{00}_2$}

\multiput(184,0)(4,0){21}{\circle*{1}}
\multiput(184,-40)(4,0){21}{\circle*{1}}
\multiput(184,-40)(0,4){11}{\circle*{1}}
\multiput(264,-40)(0,4){11}{\circle*{1}}
\put(210,-26){$G^{0m}_2$}

\multiput(0,-48)(4,0){11}{\circle*{1}}
\multiput(0,-128)(4,0){11}{\circle*{1}}
\multiput(0,-128)(0,4){21}{\circle*{1}}
\multiput(40,-128)(0,4){21}{\circle*{1}}
\put(8,-94){$G^{n0}_3$}

\multiput(48,-48)(4,0){21}{\circle*{1}}
\multiput(48,-128)(4,0){21}{\circle*{1}}
\multiput(48,-128)(0,4){21}{\circle*{1}}
\multiput(128,-128)(0,4){21}{\circle*{1}}
\put(74,-94){$G^{nm}_3$}

\multiput(136,-48)(4,0){11}{\circle*{1}}
\multiput(136,-128)(4,0){11}{\circle*{1}}
\multiput(136,-128)(0,4){21}{\circle*{1}}
\multiput(176,-128)(0,4){21}{\circle*{1}}
\put(144,-94){$G^{n0}_2$}

\multiput(184,-48)(4,0){21}{\circle*{1}}
\multiput(184,-128)(4,0){21}{\circle*{1}}
\multiput(184,-128)(0,4){21}{\circle*{1}}
\multiput(264,-128)(0,4){21}{\circle*{1}}
\put(210,-94){$G^{nm}_2$}
\end{picture}
\h\h\h\h\right\rgroup\nn
\eea
\vglue .5cm
\parindent 0mm
Here the tensor blocks are given by\parindent 8mm
\bea
G^{\h0n}
=-\h G^{\h n0}
=\h\h\frac{i\h\varepsilon^{nm}k^m}{{\bf k}^2}\h\h G^{\h\rm CS},\nn
\eea
\bea
G^{\h nm}
\h=\h\h\frac{\varepsilon^{ni}\varepsilon^{mj}k^ik^j}{{\bf k}^4}\h\h G^{\h\rm M}
+\h\h\frac{k^nk^m}{{\bf k}^4}\h\h G^{\h\parallel},\nn
\eea
\bea
G^{00}_1\h=\h\h\frac{\delta}{{\cal D}\mu_0e^2}\left\{
\frac{\mu_0e^2}{m\delta}\h\h\Pi^{\rm tot}_{\rm M}\h
+\h\left(1-\h\h\frac{\pi}{\nu_0}\h\h\Pi^{\rm tot}_{\rm CS}\right)^2\h
-\h\h\frac{\pi^2}{\nu_0^2}
\h\h\Pi^{\rm tot}_{\rm E}\Pi^{\rm tot}_{\rm M}\right\},
\eea
\bea
G^{\rm CS}_1\h=\h\h\frac{\pi}{{\cal D}\nu_0}\left\{
\Pi^{\rm tot}_{\rm CS}\h\Pi^{\rm tot}_{\rm CS}\h
-\h\Pi^{\rm tot}_{\rm E}\Pi^{\rm tot}_{\rm M}\h
-\h\h\frac{\nu_0}{\pi}\h\h\Pi^{\rm tot}_{\rm CS}\right\},
\eea
\bea
G^{\rm M}_1\h=\h\h\frac{m}{\cal D}\h\h\Pi^{\rm tot}_{\rm E}\h-\h\h
\frac{\delta{\bf k}^2}{{\cal D}\mu_0e^2c^2}\h\h\left\{
\left(1-\h\h\frac{\pi}{\nu_0}\h\h\Pi^{\rm tot}_{\rm CS}\right)^2\h
-\h\h\frac{\pi^2}{\nu_0^2}
\h\h\Pi^{\rm tot}_{\rm E}\Pi^{\rm tot}_{\rm M}\right\},
\eea
\bea
{\cal D}\h\h=\h\h
\Pi^{\rm tot}_{\rm CS}\h\Pi^{\rm tot}_{\rm CS}\h
-\h\Pi^{\rm tot}_{\rm E}\Pi^{\rm tot}_{\rm M}\h
-\h\h\frac{m\delta}{\mu_0e^2}\h\h\Pi^{\rm tot}_{\rm E}\h\h+\nn
\eea
\bea
+\h\h\frac{\delta^2{\bf k}^2}{\mu_0^2e^4c^2}
\left\{\frac{\mu_0e^2}{m\delta}\h\h\Pi^{\rm tot}_{\rm M}\h
+\h\left(1-\h\h\frac{\pi}{\nu_0}\h\h\Pi^{\rm tot}_{\rm CS}\right)^2\h
-\h\h\frac{\pi^2}{\nu_0^2}
\h\h\Pi^{\rm tot}_{\rm E}\Pi^{\rm tot}_{\rm M}\right\},\nn
\eea
and $(\nu_0/\pi)^2{\cal D}$ is the determinant of $D$. In the Coulomb gauge
$(\alpha=0)$ we obtain
\bea
G^{00}_1(k)\h=\h\h\frac
{g^{\rm E}_0+g^{\rm E}_2{\bf k}^2+g^{\rm E}_4{\bf k}^4+\cdots}
{d_0+d_2{\bf k}^2+d_4{\bf k}^4+\cdots},\nn
\eea
\bea
G^{\h0n}_1(k)\h=\h i\h\varepsilon^{nm}k^m
\left(\frac{g^{\rm CS}}{{\bf k}^2}\h\h+\h\h\frac
{g^{\rm CS}_0+g^{\rm CS}_2{\bf k}^2+g^{\rm CS}_4{\bf k}^4+\cdots}
{d_0+d_2{\bf k}^2+d_4{\bf k}^4+\cdots}\right),\nn
\eea
\bea
G^{\h nm}_1(k)\h=\h\left(\delta^{nm}\h-\h\h\frac{k^nk^m}{{\bf k}^2}\right)
\left(\frac{g^{\rm M}}{{\bf k}^2}\h\h+\h\h\frac
{g^{\rm M}_0+g^{\rm M}_2{\bf k}^2+g^{\rm M}_4{\bf k}^4+\cdots}
{d_0+d_2{\bf k}^2+d_4{\bf k}^4+\cdots}\right).\nn
\eea
The constants $g$'s are the coefficients of the momentum expansion of the
structure functions in the nominators of (36), (37) and (38), while $d$'s are
set by ${\cal D}=\sum d_{2n}{\bf k}^{2n}$.

The ratios
\bea
\frac{g_0+g_2{\bf k}^2+g_4{\bf k}^4+\cdots}
{d_0+d_2{\bf k}^2+d_4{\bf k}^4+\cdots}\nn
\eea
represent the short range interaction.

The leading contributions to the first two coefficients of the determinant
expansion are given by
\bea
d_0\h
=\h\Pi^{\rm tot}_{\rm CS}(0)\h\Pi^{\rm tot}_{\rm CS}(0)\h
-\h\h\frac{m\delta}{\mu_0e^2}\h\h\Pi^{\rm tot}_{\rm E}(0),\nn
\eea
\bea
d_2\h=\h-\h\h\frac{m\delta}{\mu_0e^2}\h\h\frac{\ell^2}{2\hbar^2}
\h\h\frac{d\h\Pi^{\rm tot}_{\rm E}(x)}{d\h x}\h\h\bigg\vert_{x=0}.\nn
\eea
Using (A.1) -- (A.5), one can show that $d_0$ and $d_2$ are the nonvanishing
positive numbers. Consequently keeping the lowest order terms one can write
\bea
\frac
{g_0+g_2{\bf k}^2+g_4{\bf k}^4+\cdots}
{d_0+d_2{\bf k}^2+d_4{\bf k}^4+\cdots}\h\h\h\h\to\h\h\h\h
\frac{g_0}{d_0+d_2{\bf k}^2}\h\h=\h\h\frac{g_0}{d_2}\h\h\h\h
\frac{1}{{\bf k}^2+{\cal M}^2},\nn
\eea
where ${\cal M}^2\h=\h d_0/d_2>0$ gives rise to the short range interaction.
Using the zero temperature values of the structure functions (see Appendix),
one can estimate the corresponding value of the mass parameter
\bea
{\cal M}^2\h=\h\h\frac{\hbar^2\mu_0e^2n_e}{m\delta}
\h\h=\h\h\frac{\hbar^2}{\lambda^2_{\rm L}},
\eea
where $\lambda_{\rm L}$ is the London penetration depth. Note, that ${\cal M}$
does not depend on $N$.

So, in the lowest order approximation $({\cal D}=d_0+d_2{\bf k}^2)$ the
spectrum of the short range interaction consists of one single mass ${\cal M}$.
According to \cite{Daemi}, the low temperature mass spectrum in
${\bf k}^4$ approximation consists of two masses ${\cal M}_+<<{\cal M}_-$ where
${\cal M}_+$ exactly coincides with ${\cal M}$ presented above.

Consider this point more carefully. Taking the case of $\nu=1$ $(N=0)$ we
present the zero temperature expression for ${\cal D}$ up to the ${\bf k}^6$
terms
\bea
\pi^2{\cal D}(x)\h=\h1\h+\h2\alpha x\h-\h\h\frac32\h\h\alpha
\left(1\h-\h\h\frac83\h\h\gamma\right)x^2\h+\h\h\frac{11}{18}\h\h\alpha
\left(1\h-\h\h\frac{81}{11}\gamma\right)x^3,
\eea
\bea
x\h=\h\h\frac{{\bf k}^2}{2\pi n_e\hbar^2},\hhh
\alpha\h=\h\pi n_e\lambda^2_{\rm L},\hhh
\gamma\h=\h\pi^2n_e^2\lambda^2_{\rm L}\ell^2_0.\nn
\eea

The analysis of \cite{Daemi} is based on the following ${\bf k}^4$
approximation
\bea
\pi^2\Delta(x)\h=\h1\h+\h2\alpha x\h-\h\h\frac32\h\h\alpha
\left(-\h\h\frac83\h\h\gamma\right)x^2\nn
\eea
and the solution to the equation $\Delta(x)=0$ leads to the conclusion about
the
existence of the second mass.

In our case the ${\bf k}^4$ approximation leads to the following factorization
\bea
{\cal D}\sim({\bf k}^2+{\cal M}^2)({\bf k}^2-a^2),\nn
\eea
where the second factor is of a tachionic type. Situation can be clarified if
the higher order terms are taken into account.
For the typical values of the
parameters we have $\gamma\sim0.0416$ and one can easily verify that the
equation
\bea
\frac{6\pi^2}{\alpha}\h\h{\cal D}'(x)\h
=\h11\left(1\h-\h\h\frac{81}{11}\gamma\right)x^2\h
-\h18\left(1\h-\h\h\frac83\h\h\gamma\right)x\h+12\h=\h0\nn
\eea
has no real solution, i.e. ${\cal D}(x)$ exhibits no extrema. This fact
together with ${\cal D}'(0)>0$ implyies that ${\cal D}'(x)>0$ for all $x$.
Moreover, we have ${\cal D}(0)=1$ which means that defining equation
${\cal D}(x)=0$ has a single real solution $x=x_0<0$. It is not difficult to
guess that $x_0$ tends to $-{\cal M}^2$ when we go to the lower approximation
$\pi^2{\cal D}=1+2\alpha x$. In other words, ${\bf k}^6$ order corrections do
not affect the lower order result: the short range interaction spectrum derived
from (40) consists of a single mass defined by (39). The same time it is
obvious, that the presented assertion depends on the approximation used and
can be changed in higher orders.

Constant parameters, corresponding to the long range components are given by
\bea
g^{\rm CS}\h
=\h\h\frac{\pi}{\nu_0d_0}\h\h\{\Pi^{\rm tot}_{\rm CS}(0)
\h\Pi^{\rm tot}_{\rm CS}(0)\h
-\h\Pi^{\rm tot}_{\rm E}(0)\h\Pi^{\rm tot}_{\rm M}(0)\h
-\h\h\frac{\nu_0}{\pi}\h\h\Pi^{\rm tot}_{\rm CS}(0)\},\nn
\eea
\bea
g^{\rm M}\h=\h\h\frac{m}{d_0}\h\h\Pi^{\rm tot}_{\rm E}(0),\nn
\eea
and using (A.6) -- (A.8) one can show that they vanish at $T=0$ for any
$\nu_0=N+1$.

So, at finite temperatures the Coulomb forces carry the finite range character,
while the charge\h-\h current and the current\h-\h current interactions include
the short as well as the long range parts. When the temperature is decreased,
integer number of the Landau levels are kept to be completely filled, and we
see that the strength of the long range interaction decreases until vanishes at
$T=0^{\circ}K$, where the Maxwell field completely acquires the finite range
character.

\vspace*{15mm}
\parindent 0mm
\def\theequation{A.\arabic{equation}}
\setcounter{equation}{0}
{\bf Appendix}
\parindent 8mm
\vspace*{5mm}

Structure functions for a one type fermion system can be presented in the
following analytical form
\bea
\Pi_{\rm E}(x)=-\h\h\frac{h}{2\pi}\h\h\h I(x)
-\h\h\frac{1}{4\pi}\h\h\h\h e^{-\h x}\h S_3(x,x)
+\h\h\frac{h}{2\pi}\h\h\h\h e^{-\h x}\h\h\frac{\partial}{\partial x}\h\h
x\h\h\frac{\partial}{\partial x}\h\h\h S_0(x,y)
\h\bigg\vert_{y=x},
\eea
\bea
\Pi_{\rm CS}(x)
=\h\h\frac{1}{4\pi}\h\h\h\h e^{-\h x}\left[\h\frac{\partial}{\partial x}
\h\h\h S_1(x,y)-\h\h\frac{\partial}{\partial x}\h\h\h S_2(x,y)
+S_2(x,y)\h\right]\bigg\vert_{y=x},
\eea
\bea
\Pi_{\rm M}(x)
=-\h\h\frac{1}{2\pi}\h\h\h\h e^{-\h x}\h\h\frac{\partial^2}{\partial x\partial
y}
\h\h\h S_0(x,y)\h\bigg\vert_{y=x},
\eea
where $I(x)$ and $S_a(x,y)$ are given by
\bea
I(x)
=e^{-\h x}\sum_{n=0}^{\infty}\sum_{\alpha=1}^{\infty}
\frac{n\h!}{(n+\alpha)\h!}
\h\h\h\h\alpha^2\h\h\Theta_0(n,\alpha)\h\h
x^{\alpha-1}L_n^{\alpha}(x)\h L_n^{\alpha}(x),
\eea
\bea
S_a(x,y)
=\sum_{n=0}^{\infty}\Theta_a(n)L_n(x)L_n(y)+\nn
\eea
\bea
+\sum_{n=0}^{\infty}\sum_{\alpha=1}^{\infty}
\frac{n\h!}{(n+\alpha)\h!}\h\h\Theta_a(n,\alpha)\h(x^{\alpha}+y^{\alpha})\h
L_n^{\alpha}(x)\h L_n^{\alpha}(y).
\eea
Here $L^{\alpha}_n(x)$ are the generalized Laguerre polynomials, while
$\Theta_a$ are the temperature dependent constants
\bea
\alpha\h\Theta_a(n,\alpha)\h
=\h\theta_a(n)\h-\h\theta_a(n+\alpha),
\hh\hh\Theta_a(n)\h=\h\lim_{\alpha\to0}\Theta_a(n,\alpha),\nn
\eea
\bea
\theta_0(n)=\frac{\rho_n}{\omega_n},
\hhh\theta_1(n)=\sigma\h\h\frac{\rho_n}{\omega_n}+\varepsilon\h\bar\rho_n,
\hhh\theta_2(n)
=\sigma\h\h\frac{\rho_{n+1}}{\omega_{n+1}}-\varepsilon\h\bar\rho_{n+1},\nn
\eea
\bea
\theta_3(n)=\h\h\frac12\h\left(\omega_n+\h\frac{1}{\omega_n}\right)\rho_n+
\h\h\frac12\h\left(\omega_{n+1}+\h\frac{1}{\omega_{n+1}}\right)\rho_{n+1}+
\sigma\varepsilon\h(\bar\rho_n-\h\bar\rho_{n+1}),\nn
\eea
\bea
\rho_n\equiv\rho^+_n(\mu)+\h\rho^-_n(\mu),\hhh\hhh
\bar\rho_n\equiv\rho^+_n(\mu)-\h\rho^-_n(\mu).\nn
\eea

Structure functions which include the contributions from both sorts of fermions
are denoted as $\Pi^{\rm tot}_{\rm E}(x)$, $\Pi^{\rm tot}_{\rm CS}(x)$ and
$\Pi^{\rm tot}_{\rm M}(x)$. For our purposes we consider the case of $T=0$ and
$\bar\alpha=0$. If the applied magnetic field $B^{\rm ext}$ does not exceed the
critical one, then at $T=0$ it is completely expelled from the sample
$(B^{\rm b}=0)$, and according to (24) and (25) we have
$\nu_1=\nu_2=\nu=\nu_0=N+1$ where $N$ is any nonnegative integer. Consequently,
only the nonvanishing terms in (20) and (21) in the zero temperature limit
$(\beta=\infty)$ are
\bea
\rho^+_0(\mu_1)
\h=\h\cdots\h=\h\rho^+_{N-1}(\mu_1)\h=\h\rho^+_N(\mu_1)\h=\h1,\nn
\eea
\bea
\rho^+_1(\mu_2)
\h=\h\cdots\h=\h\rho^+_N(\mu_2)\h=\h\rho^+_{N+1}(\mu_2)\h=\h1,\nn
\eea
while all others, including the antiparticle contributions vanish exactly.
These values of the Fermi distribution functions lead to the following zero
temperature expressions
\bea
\Pi^{\rm tot}_{\rm E}(x)\h=\h\h\frac{1}{2\pi}\sum_{n=0}^N\left\{
-4x\h+\h3x^2(2n\h+\h1)\h-\h\h\frac{x^3}{9}\h\h(30n^2\h+\h30n\h+\h11)\right\},
\eea
\bea
\Pi^{\rm tot}_{\rm CS}(x)\h=\h\h\frac{1}{\pi}\sum_{n=0}^N\left\{
1\h-\h\h\frac{3x}{2}\h\h(2n\h+\h1)\h
+\h\h\frac{x^2}{12}\h\h(30n^2\h+\h30n\h+\h11)\right\}+\nn
\eea
\bea
+\h\h\frac{x^3}{72\pi}\sum_{n=0}^N(70n^3\h+\h105n^2\h+\h85n\h+\h25),
\eea
\bea
\Pi^{\rm tot}_{\rm M}(x)\h=\h\h\frac{1}{\pi}\sum_{n=0}^N\left\{
2n\h+\h1\h-\h\h\frac{3x}{2}\h\h(2n^2\h+\h2n\h+\h1)\right\}+\nn
\eea
\bea
+\h\h\frac{x^2}{12\pi}\sum_{n=0}^N(30n^3\h+\h30n^2\h+\h32n\h+\h11)\h+\nn
\eea
\bea
+\h\h\frac{x^3}{72\pi}\sum_{n=0}^N(35n^4\h+\h70n^3\h+\h120n^2\h+\h85n\h+\h25),
\eea
where the structure functions are expanded up to $x^3$ i.e. up to ${\bf k}^6$
terms. Note, that the closed expressions for the nonrelativistic structure
functions are presented in \cite{Eliashvili}.

   \newcommand{\r}{\circle*{2}}
   \newcommand{\rr}{\circle*{5}}
   \newpage

   \begin{center}
   \setlength{\unitlength}{.1mm}
   \begin{picture}(1520,900)

   \multiput(100,60)(20,0){67}{\rr}    \multiput(160,60)(200,0){7}{\circle{10}}
   \multiput(160,0)(0,20){46}{\rr}

   \put(349,10){1}
   \put(550,10){2}
   \put(751,10){3}
   \put(951,10){4}
   \put(1151,10){5}
   \put(1350,10){$\nu$}

   \put(1250,464){$_{({\cal F}_1\hspace*{.2mm}+\h{\cal F}_2)\h/\h2}$}
   \put(1261,540){$_{{\cal F}_1}$}
   \put(1261,384){$_{{\cal F}_2}$}

   \multiput(310,155)(10,1){5}{\rr}	\multiput(360,760)(-5,10){3}{\rr}
   \put(344,790){\rr}			\put(336,800){\rr}
   \put(326,809){\rr} 			\put(316,818){\rr}

   \put(357,471){\rr} 			\put(350,480){\rr}
   \put(342,490){\rr}			\put(332,498){\rr}
   \put(322,506){\rr}			\put(310,512){\rr}

   \put(363,171){\rr}			\put(557,681){\rr}
   \put(363,471){\rr}			\put(555,469){\rr}

   \put(360,160){\rr}	  \put(360,760){\rr}	 \put(360,460){\rr}
   \put(370,182){\rr}	  \put(370,759){\rr}	 \put(370,480){\rr}
   \put(380,192){\rr}	  \put(380,758){\rr}	 \put(380,490){\rr}
   \put(390,200){\rr}	  \put(390,757){\rr}	 \put(390,496){\rr}
   \put(400,206){\rr}	  \put(400,756){\rr}	 \put(400,502){\rr}
   \put(410,212){\rr}	  \put(410,754){\rr}	 \put(410,505){\rr}
   \put(420,217){\rr}	  \put(420,752){\rr}	 \put(420,508){\rr}
   \put(430,221){\rr}	  \put(430,750){\rr}	 \put(430,510){\rr}
   \put(440,225){\rr}	  \put(440,747){\rr}	 \put(440,511){\rr}
   \put(450,229){\rr}	  \put(450,745){\rr}	 \put(450,511){\rr}
   \put(460,232){\rr}	  \put(460,742){\rr}	 \put(460,511){\rr}
   \put(470,235){\rr}	  \put(470,739){\rr}	 \put(470,510){\rr}
   \put(480,237){\rr}	  \put(480,735){\rr}	 \put(480,509){\rr}
   \put(490,240){\rr}	  \put(490,731){\rr}	 \put(490,506){\rr}
   \put(500,242){\rr}	  \put(500,727){\rr}	 \put(500,504){\rr}
   \put(510,244){\rr}	  \put(510,722){\rr}	 \put(510,500){\rr}
   \put(520,245){\rr}	  \put(520,716){\rr}	 \put(520,496){\rr}
   \put(530,247){\rr}	  \put(530,710){\rr}	 \put(530,490){\rr}
   \put(540,248){\rr}	  \put(540,702){\rr}	 \put(540,484){\rr}
   \put(550,249){\rr}	  \put(550,691){\rr}	 \put(547,477){\rr}

   \put(560,250){\rr}     \put(560,670){\rr}	 \put(560,460){\rr}

   \put(565,258){\rr}	  \put(570,670){\rr}	 \put(565,468){\rr}
   \put(572,266){\rr} 	  \put(580,669){\rr}	 \put(572,475){\rr}
   \put(580,273){\rr}	  \put(590,669){\rr}	 \put(580,482){\rr}
   \put(590,279){\rr}	  \put(600,668){\rr}	 \put(590,487){\rr}
   \put(600,284){\rr}	  \put(610,668){\rr}	 \put(600,491){\rr}
   \put(610,288){\rr}	  \put(620,667){\rr}	 \put(610,495){\rr}
   \put(620,291){\rr}	  \put(630,666){\rr}	 \put(620,497){\rr}
   \put(630,293){\rr}	  \put(640,665){\rr}	 \put(630,498){\rr}
   \put(640,296){\rr}	  \put(650,663){\rr}	 \put(640,499){\rr}
   \put(650,298){\rr}	  \put(660,661){\rr}	 \put(650,499){\rr}
   \put(660,301){\rr}	  \put(670,659){\rr}	 \put(660,499){\rr}
   \put(670,302){\rr}	  \put(680,656){\rr}	 \put(670,498){\rr}
   \put(680,303){\rr}	  \put(690,654){\rr}	 \put(680,497){\rr}
   \put(690,304){\rr}	  \put(700,651){\rr}	 \put(690,496){\rr}
   \put(700,305){\rr}	  \put(710,648){\rr}	 \put(700,493){\rr}
   \put(710,306){\rr}	  \put(720,644){\rr}	 \put(710,490){\rr}
   \put(720,308){\rr}	  \put(730,639){\rr}	 \put(720,486){\rr}
   \put(730,309){\rr}	  \put(740,633){\rr}	 \put(730,481){\rr}
   \put(740,309){\rr}	  \put(748,626){\rr}	 \put(740,475){\rr}
   \put(750,310){\rr}	  \put(755,618){\rr}	 \put(750,468){\rr}

   \put(760,310){\rr}	  \put(760,610){\rr}	 \put(760,460){\rr}

   \put(765,316){\rr}	  \put(770,610){\rr}	 \put(765,466){\rr}
   \put(772,322){\rr}	  \put(780,609){\rr}	 \put(772,471){\rr}
   \put(780,327){\rr}	  \put(790,609){\rr}	 \put(780,476){\rr}
   \put(790,330){\rr}	  \put(800,608){\rr}	 \put(790,479){\rr}
   \put(800,334){\rr}	  \put(810,608){\rr}	 \put(800,482){\rr}
   \put(810,336){\rr}	  \put(820,606){\rr}	 \put(810,483){\rr}
   \put(820,338){\rr}	  \put(830,606){\rr}	 \put(820,484){\rr}
   \put(830,339){\rr}	  \put(840,605){\rr}	 \put(830,485){\rr}
   \put(840,340){\rr}	  \put(850,604){\rr}	 \put(840,485){\rr}
   \put(850,342){\rr}	  \put(860,603){\rr}	 \put(850,486){\rr}
   \put(860,343){\rr}	  \put(870,602){\rr}	 \put(860,486){\rr}
   \put(870,344){\rr}	  \put(880,600){\rr}	 \put(870,485){\rr}
   \put(880,345){\rr}	  \put(890,599){\rr}	 \put(880,485){\rr}
   \put(890,346){\rr}	  \put(900,598){\rr}	 \put(890,484){\rr}
   \put(900,347){\rr}	  \put(910,596){\rr}	 \put(900,483){\rr}
   \put(910,348){\rr}	  \put(920,593){\rr}	 \put(910,481){\rr}
   \put(920,348){\rr}	  \put(930,590){\rr}	 \put(920,478){\rr}
   \put(930,349){\rr}	  \put(939,586){\rr}	 \put(930,475){\rr}
   \put(940,349){\rr}	  \put(947,582){\rr}	 \put(940,471){\rr}
   \put(950,350){\rr}	  \put(955,576){\rr}	 \put(950,466){\rr}

   \put(960,350){\rr}	  \put(960,570){\rr}	 \put(960,460){\rr}

   \put(970,355){\rr}	  \put(970,570){\rr}	 \put(970,464){\rr}
   \put(980,358){\rr}	  \put(980,570){\rr}	 \put(980,468){\rr}
   \put(990,360){\rr}	  \put(990,570){\rr}	 \put(990,470){\rr}
   \put(1000,363){\rr}	  \put(1000,570){\rr}	 \put(1000,472){\rr}
   \put(1010,364){\rr}	  \put(1010,569){\rr}	 \put(1010,474){\rr}
   \put(1020,365){\rr}	  \put(1020,569){\rr}	 \put(1020,475){\rr}
   \put(1030,366){\rr}	  \put(1030,569){\rr}	 \put(1030,476){\rr}
   \put(1040,368){\rr}	  \put(1040,569){\rr}	 \put(1040,477){\rr}
   \put(1050,369){\rr}	  \put(1050,567){\rr}	 \put(1050,477){\rr}
   \put(1060,370){\rr}	  \put(1060,567){\rr}	 \put(1060,477){\rr}
   \put(1070,370){\rr}	  \put(1070,566){\rr}	 \put(1070,477){\rr}
   \put(1080,371){\rr}	  \put(1080,566){\rr}	 \put(1080,477){\rr}
   \put(1090,371){\rr}	  \put(1090,565){\rr}	 \put(1090,476){\rr}
   \put(1100,372){\rr}	  \put(1100,564){\rr}	 \put(1100,475){\rr}
   \put(1110,372){\rr}	  \put(1110,562){\rr}	 \put(1110,474){\rr}
   \put(1120,373){\rr}	  \put(1120,559){\rr}	 \put(1120,472){\rr}
   \put(1130,373){\rr}	  \put(1130,557){\rr}	 \put(1130,470){\rr}
   \put(1140,374){\rr}	  \put(1140,555){\rr}	 \put(1140,468){\rr}
   \put(1150,374){\rr}	  \put(1150,551){\rr}	 \put(1150,464){\rr}

   \put(1160,374){\rr}	  \put(1160,546){\rr}	 \put(1160,460){\rr}

   \put(1170,378){\rr}	 \put(1170,545){\rr}	 \put(1170,464){\rr}
   \put(1180,381){\rr}	 \put(1180,545){\rr}	 \put(1180,466){\rr}
   \put(1190,384){\rr}	 \put(1190,544){\rr}	 \put(1190,467){\rr}
   \put(1200,386){\rr}	 \put(1200,543){\rr}	 \put(1200,468){\rr}

   \end{picture}
   \end{center}
   \begin{center}
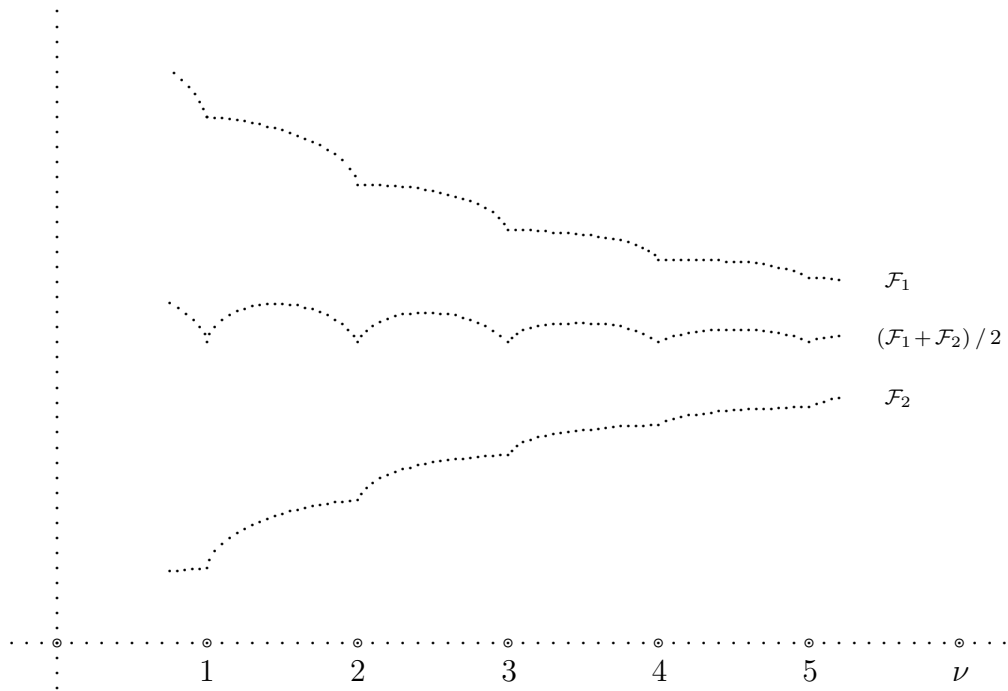
   Figure 1. Free energies $(conventional\h\h units)$ versus $\nu$
             for $\bar\alpha=0$ and $T=50^{\h\circ}K$.
   \end{center}
   \vglue 16mm
   \begin{center}
   \setlength{\unitlength}{.25mm}
   \begin{picture}(608,360)

   % origin = ( 304 , 184 )

   \multiput(24,24)(8,0){71}{\r}	\multiput(64,24)(80,0){7}{\circle{4}}
   \multiput(64,0)(0,8){46}{\r}         \multiput(64,104)(0,80){4}{\circle{4}}

   \put(40,342){$M$}			\put(24,259){$+\h60$}
   \put(48,179){$0$}	        	\put(24,99){$-\h60$}

   \put(116,4){$-\h120$}		\put(200,4){$-\h60$}
   \put(300,4){$0$}			\put(361,4){$+\h60$}
   \put(437,4){$+\h120$}		\put(536,4){$B^{\rm b}$}

   \put(541,37){$0\h^{\circ}K$}
   \put(536,88){$40\h^{\circ}K$}
   \put(536,141){$80\h^{\circ}K$}

   % temperature = 0 grad

   \multiput(72,336)(4,0){58}{\r}	\multiput(308,32)(4,0){69}{\r}
   \multiput(304,334)(0,-4){15}{\r}	\multiput(304,34)(0,4){15}{\r}

   % temperature = 40 grad

   \multiput(71,288)(4,0){8}{\r}	\multiput(509,80)(4,0){19}{\r}
   \multiput(103,287)(4,0){10}{\r}	\multiput(469,81)(4,0){10}{\r}
   \multiput(143,286)(4,0){8}{\r}	\multiput(437,82)(4,0){8}{\r}
   \multiput(175,285)(4,0){7}{\r}	\multiput(409,83)(4,0){7}{\r}
   \multiput(203,284)(4,0){5}{\r}	\multiput(389,84)(4,0){5}{\r}
   \multiput(223,283)(4,0){4}{\r}	\multiput(373,85)(4,0){4}{\r}
   \multiput(239,282)(4,0){3}{\r}	\multiput(361,86)(4,0){3}{\r}
   \multiput(251,281)(4,0){3}{\r}	\multiput(349,87)(4,0){3}{\r}

   \put(263,280){\r}			\put(345,88){\r}
   \put(267,280){\r}			\put(341,88){\r}
   \put(271,279){\r}			\put(337,89){\r}
   \put(275,279){\r}			\put(333,89){\r}
   \put(279,278){\r}			\put(329,90){\r}
   \put(283,277){\r}			\put(325,91){\r}
   \put(287,276){\r}			\put(321,92){\r}
   \put(291,275){\r}			\put(317,93){\r}
   \put(295,274){\r}			\put(313,94){\r}
   \put(299,272){\r}			\put(309,96){\r}
   \put(302,269){\r}			\put(306,99){\r}
   \put(303,265){\r}			\put(305,103){\r}

   \multiput(304,108)(0,4){39}{\r}

   % temperature = 80 grad

   \multiput(73,235)(4,0){2}{\r}	\multiput(531,133)(4,0){13}{\r}
   \multiput(81,234)(4,0){6}{\r}	\multiput(507,134)(4,0){6}{\r}
   \multiput(105,233)(4,0){6}{\r}	\multiput(483,135)(4,0){6}{\r}
   \multiput(129,232)(4,0){4}{\r}	\multiput(467,136)(4,0){4}{\r}
   \multiput(145,231)(4,0){5}{\r}	\multiput(447,137)(4,0){5}{\r}
   \multiput(165,230)(4,0){3}{\r}	\multiput(435,138)(4,0){3}{\r}
   \multiput(177,229)(4,0){4}{\r}	\multiput(419,139)(4,0){4}{\r}
   \multiput(193,228)(4,0){3}{\r}	\multiput(407,140)(4,0){3}{\r}
   \multiput(205,227)(4,0){3}{\r}	\multiput(395,141)(4,0){3}{\r}

   \put(217,226){\r}			\put(391,142){\r}
   \put(221,226){\r}			\put(387,142){\r}
   \put(225,225){\r}			\put(383,143){\r}
   \put(229,225){\r}			\put(379,143){\r}
   \put(233,224){\r}			\put(375,144){\r}
   \put(237,224){\r}			\put(371,144){\r}
   \put(241,223){\r}			\put(367,145){\r}
   \put(245,223){\r}			\put(363,145){\r}
   \put(249,222){\r}			\put(359,146){\r}
   \put(253,221){\r}			\put(355,147){\r}
   \put(257,221){\r}			\put(351,147){\r}
   \put(261,220){\r}			\put(347,148){\r}
   \put(265,219){\r}			\put(343,149){\r}
   \put(269,218){\r}			\put(339,150){\r}
   \put(273,216){\r}			\put(335,152){\r}
   \put(277,214){\r}			\put(331,154){\r}
   \put(281,212){\r}			\put(327,156){\r}
   \put(285,209){\r}			\put(323,159){\r}
   \put(289,206){\r}			\put(319,162){\r}
   \put(292,203){\r}			\put(316,165){\r}
   \put(295,199){\r}			\put(313,169){\r}
   \put(298,195){\r}			\put(310,173){\r}
   \put(301,190){\r}			\put(307,178){\r}

   \end{picture}
   \end{center}
   \begin{center}
   Figure 2. $M$ $[\h Gauss]$ versus $B^{\rm b}$ $[\h Gauss]$
             for $\bar\alpha=0$.
   \end{center}

   \newpage

   \begin{center}
   \setlength{\unitlength}{.25mm}
   \begin{picture}(608,360)

   % origin = ( 304 , 184 )

   \multiput(24,24)(8,0){71}{\r}	\multiput(64,24)(80,0){7}{\circle{4}}
   \multiput(64,0)(0,8){46}{\r}		\multiput(64,104)(0,80){4}{\circle{4}}

   \put(36,342){$B^{\rm b}$}
   \put(24,259){$+\h60$}
   \put(48,179){$0$}
   \put(24,99){$-\h60$}

   \put(116,4){$-\h120$}
   \put(200,4){$-\h60$}
   \put(300,4){$0$}
   \put(361,4){$+\h60$}
   \put(437,4){$+\h120$}
   \put(532,4){$B^{\rm ext}$}

   \put(552,266){$0^{\h\circ}K$}
   \put(544,322){$40^{\h\circ}K$}
   \put(462,342){$80^{\h\circ}K$}

   % temperature = 0 grad

   \multiput(404,184)(5,0){11}{\r}	\multiput(204,184)(-5,0){11}{\r}
   \put(459,185){\circle*{2}}		\put(149,183){\r}
   \multiput(464,187)(4,4){21}{\r}	\multiput(144,181)(-4,-4){19}{\r}

   % temperature = 40 grad

   \multiput(339,184)(5,0){9}{\r}	\multiput(269,184)(-5,0){9}{\r}
   \put(384,185){\r}			\put(224,183){\r}
   \put(389,186){\r}			\put(219,182){\r}
   \put(394,188){\r}			\put(214,180){\r}
   \put(399,190){\r}			\put(209,178){\r}
   \put(404,192){\r}			\put(204,176){\r}
   \put(408,195){\r}			\put(200,173){\r}
   \put(412,199){\r}			\put(196,169){\r}
   \multiput(416,203)(4,4){31}{\r}	\multiput(192,165)(-4,-4){31}{\r}

   % temperature = 80 grad

   \put(304,184){\r}
   \multiput(309,185)(5,1){3}{\r}  	\multiput(299,183)(-5,-1){3}{\r}
   \put(324,189){\r}                	\put(284,179){\r}
   \put(329,191){\r}                	\put(279,177){\r}
   \put(334,193){\r}		 	\put(274,175){\r}
   \put(339,196){\r}		 	\put(269,172){\r}
   \put(344,199){\r}			\put(264,169){\r}
   \multiput(348,202)(4,4){34}{\r}	\multiput(260,166)(-4,-4){34}{\r}

   \end{picture}
   \end{center}
   \begin{center}
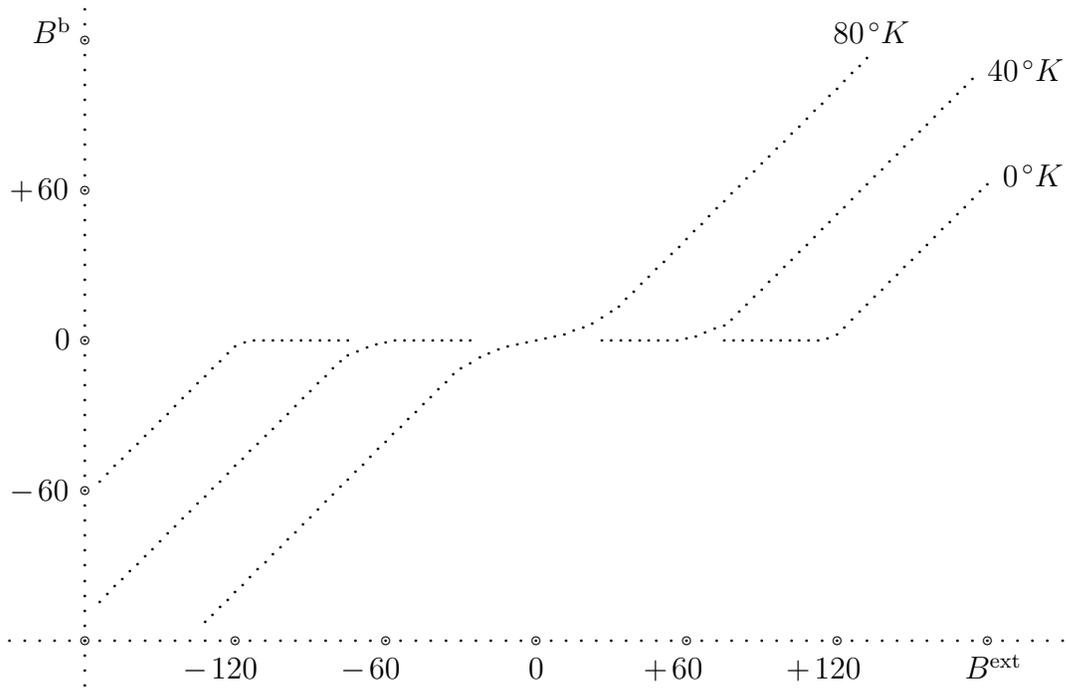
   Figure 3. $B^{\rm b}$ $[\h Gauss]$ versus $B^{\rm ext}$ $[\h Gauss]$
             for $\bar\alpha=0$.
   \end{center}
   \vglue 16mm
   \begin{center}
   \setlength{\unitlength}{.25mm}
   \begin{picture}(608,360)

   % origin = ( 64 , 24 )

   \multiput(24,24)(8,0){71}{\r}	\multiput(64,24)(80,0){7}{\circle{4}}
   \multiput(64,0)(0,8){46}{\r}		\multiput(64,104)(0,80){4}{\circle{4}}

   \put(34,342){$B^{\rm cr}$}
   \put(29,259){120}
   \put(38,179){80}
   \put(38,99){40}

   \put(136,4){20}
   \put(215,4){40}
   \put(296,4){60}
   \put(376,4){80}
   \put(451,4){100}
   \put(538,4){$T$}

   \put(68,254){\r}	\put(72,251){\r}	\put(76,248){\r}
   \put(80,245){\r}	\put(84,242){\r}	\put(88,239){\r}
   \put(92,236){\r}	\put(96,233){\r}	\put(100,230){\r}
   \put(104,227){\r}	\put(108,225){\r}	\put(112,222){\r}
   \put(116,219){\r}	\put(120,216){\r}	\put(124,214){\r}
   \put(128,211){\r}	\put(132,208){\r}	\put(136,205){\r}
   \put(140,203){\r}	\put(144,200){\r}	\put(148,197){\r}
   \put(152,195){\r}	\put(156,192){\r}	\put(160,189){\r}
   \put(164,187){\r}	\put(168,184){\r}	\put(172,181){\r}
   \put(176,179){\r}	\put(180,176){\r}	\put(184,173){\r}
   \put(188,171){\r}	\put(192,168){\r}	\put(196,165){\r}
   \put(200,163){\r}	\put(204,160){\r}	\put(208,158){\r}
   \put(212,155){\r}	\put(216,153){\r}	\put(220,150){\r}
   \put(224,148){\r}	\put(228,145){\r}	\put(232,142){\r}
   \put(236,140){\r}	\put(240,137){\r}	\put(244,135){\r}
   \put(248,132){\r}	\put(252,130){\r}	\put(256,127){\r}
   \put(260,125){\r}	\put(264,122){\r}	\put(268,119){\r}
   \put(272,117){\r}	\put(276,114){\r}	\put(280,112){\r}
   \put(284,109){\r}	\put(288,107){\r}	\put(292,104){\r}
   \put(296,102){\r}	\put(300,99){\r}	\put(304,97){\r}
   \put(308,94){\r}	\put(312,92){\r}	\put(316,89){\r}
   \put(320,87){\r}	\put(324,84){\r}	\put(328,82){\r}
   \put(332,79){\r}	\put(336,77){\r}	\put(340,74){\r}
   \put(344,72){\r}	\put(348,69){\r}	\put(352,67){\r}
   \put(356,64){\r}	\put(360,62){\r}	\put(364,60){\r}
   \put(368,57){\r}	\put(372,55){\r}	\put(376,52){\r}
   \put(380,50){\r}	\put(384,47){\r}	\put(388,45){\r}
   \put(392,42){\r}	\put(396,40){\r}	\put(400,37){\r}
   \put(404,35){\r}	\put(408,33){\r}	\put(412,30){\r}
   \put(416,28){\r}	\put(420,26){\r}
   \end{picture}
   \end{center}
   \begin{center}
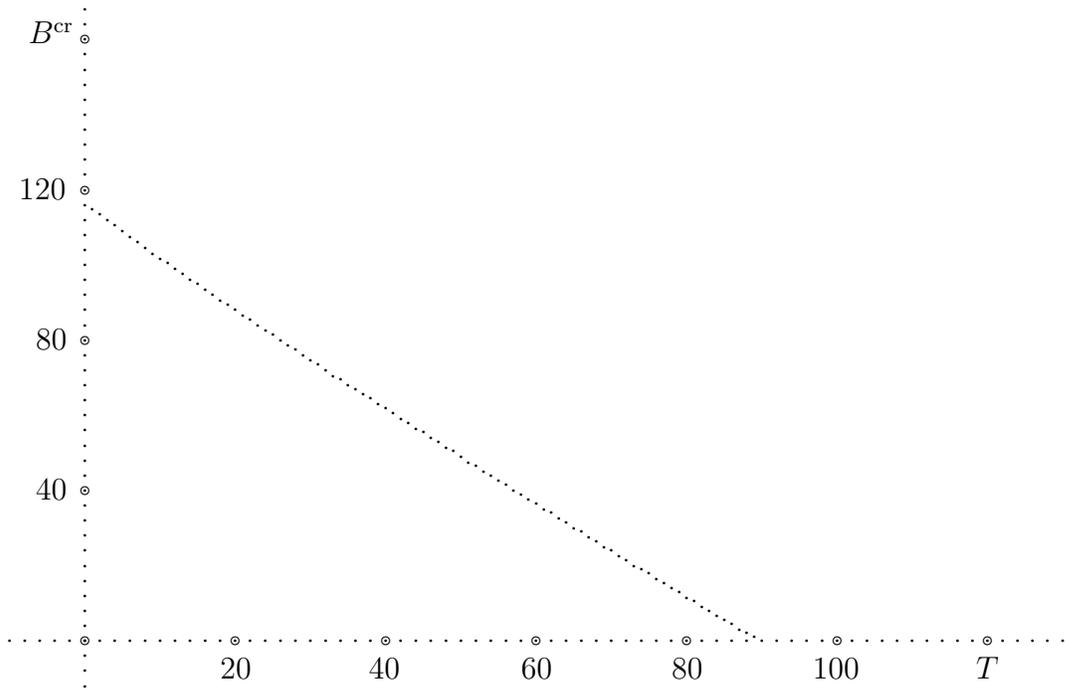
   Figure 4. Cricital magnetic field $[\h Gauss]$ versus
             temperature $[^{\circ}K]$ for $\bar\alpha=0$.
   \end{center}

   \newpage

   \begin{center}
   \setlength{\unitlength}{.25mm}
   \begin{picture}(608,360)

   % origin = ( 304 , 184 )

   \multiput(24,24)(8,0){71}{\r}	\multiput(64,24)(80,0){7}{\circle{4}}
   \multiput(64,0)(0,8){46}{\r}		\multiput(64,104)(0,80){4}{\circle{4}}

   \put(40,342){$M$}		\put(48,179){$0$}
   \put(24,259){$+\h60$}	\put(24,99){$-\h60$}

   \put(116,4){$-\h120$}   	\put(200,4){$-\h60$}
   \put(300,4){$0$}   		\put(361,4){$+\h60$}
   \put(437,4){$+\h120$}   	\put(536,4){$B^{\rm b}$}

   \multiput(72,336)(4,0){48}{\r}           \put(263,335){\r}
   \multiput(264,332)(0,-4){37}{\r}         \put(265,185){\r}
   \multiput(268,184)(4,0){19}{\r}          \put(343,183){\r}
   \multiput(344,180)(0,-4){37}{\r}         \put(345,33){\r}
   \multiput(348,32)(4,0){59}{\r}

   \end{picture}
   \end{center}
   \begin{center}
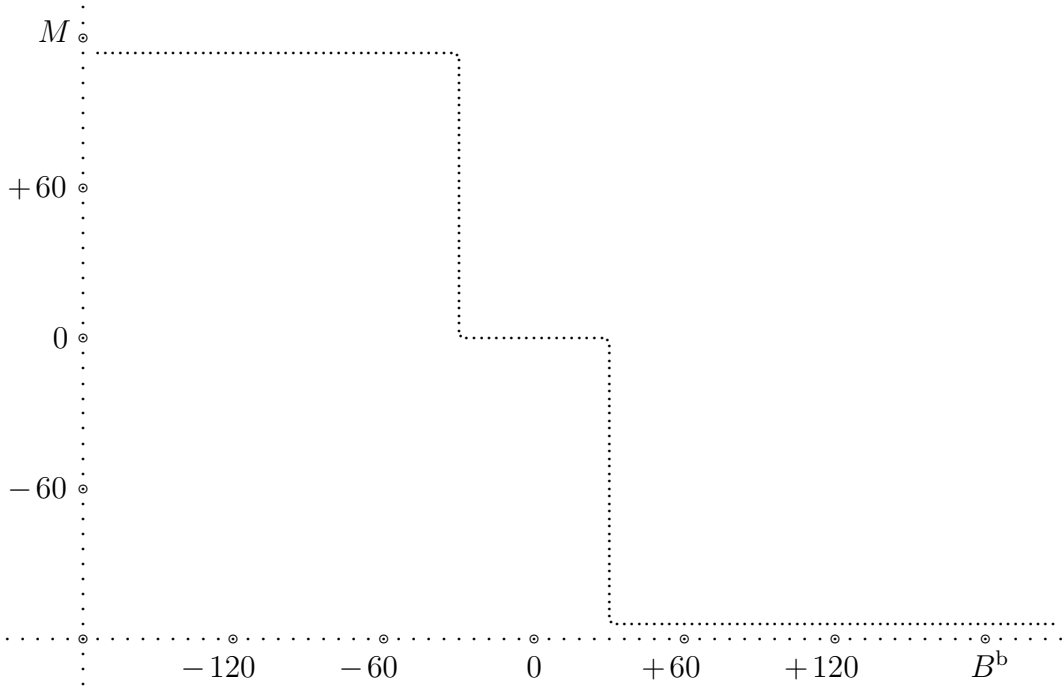
   Figure 5. $M$ $[\h Gauss]$ versus $B^{\rm b}$ $[\h Gauss]$
          for $(\pi n_e\hbar/e)\cdot\bar\alpha=30\h\h Gauss$
          and $T=0\h^{\circ}K$.
   \end{center}
   \vglue 16mm
   \begin{center}
   \setlength{\unitlength}{.25mm}
   \begin{picture}(608,360)

   % origin = ( 304 , 184 )

   \multiput(24,24)(8,0){71}{\r}	\multiput(64,24)(80,0){7}{\circle{4}}
   \multiput(64,0)(0,8){46}{\r}		\multiput(64,104)(0,80){4}{\circle{4}}

   \put(40,342){$M$}   		\put(24,259){$+\h60$}
   \put(48,179){$0$}   		\put(24,99){$-\h60$}

   \put(116,4){$-\h120$}	\put(200,4){$-\h60$}
   \put(300,4){$0$}		\put(361,4){$+\h60$}
   \put(437,4){$+\h120$}	\put(536,4){$B^{\rm b}$}

   \multiput(304,184)(4,0){2}{\r}	   \multiput(304,184)(-4,0){2}{\r}
   \multiput(312,183)(4,0){2}{\r}	   \multiput(296,185)(-4,0){2}{\r}
   \multiput(320,182)(4,0){2}{\r}	   \multiput(288,186)(-4,0){2}{\r}
   \put(328,181){\r}			   \put(280,187){\r}
   \put(332,180){\r}			   \put(276,188){\r}
   \put(336,179){\r}			   \put(272,189){\r}
   \put(340,177){\r}			   \put(268,191){\r}
   \put(343,174){\r}			   \put(265,194){\r}
   \multiput(344,170)(0,-4){18}{\r}	   \multiput(264,198)(0,4){18}{\r}
   \put(345,98){\r}			   \put(263,270){\r}
   \put(346,94){\r}			   \put(262,274){\r}
   \put(348,91){\r}			   \put(260,277){\r}
   \put(352,89){\r}			   \put(256,279){\r}
   \put(356,88){\r}			   \put(252,280){\r}
   \put(360,87){\r}			   \put(248,281){\r}
   \multiput(364,86)(4,0){2}{\r}	   \multiput(244,282)(-4,0){2}{\r}
   \multiput(372,85)(4,0){3}{\r}	   \multiput(236,283)(-4,0){3}{\r}
   \multiput(384,84)(4,0){4}{\r}	   \multiput(224,284)(-4,0){4}{\r}
   \multiput(400,83)(4,0){4}{\r}	   \multiput(208,285)(-4,0){4}{\r}
   \multiput(416,82)(4,0){6}{\r}	   \multiput(192,286)(-4,0){6}{\r}
   \multiput(440,81)(4,0){8}{\r}	   \multiput(168,287)(-4,0){8}{\r}
   \multiput(472,80)(4,0){10}{\r}	   \multiput(136,288)(-4,0){10}{\r}
   \multiput(512,79)(4,0){18}{\r}	   \multiput(96,289)(-4,0){7}{\r}

   \end{picture}
   \end{center}
   \begin{center}
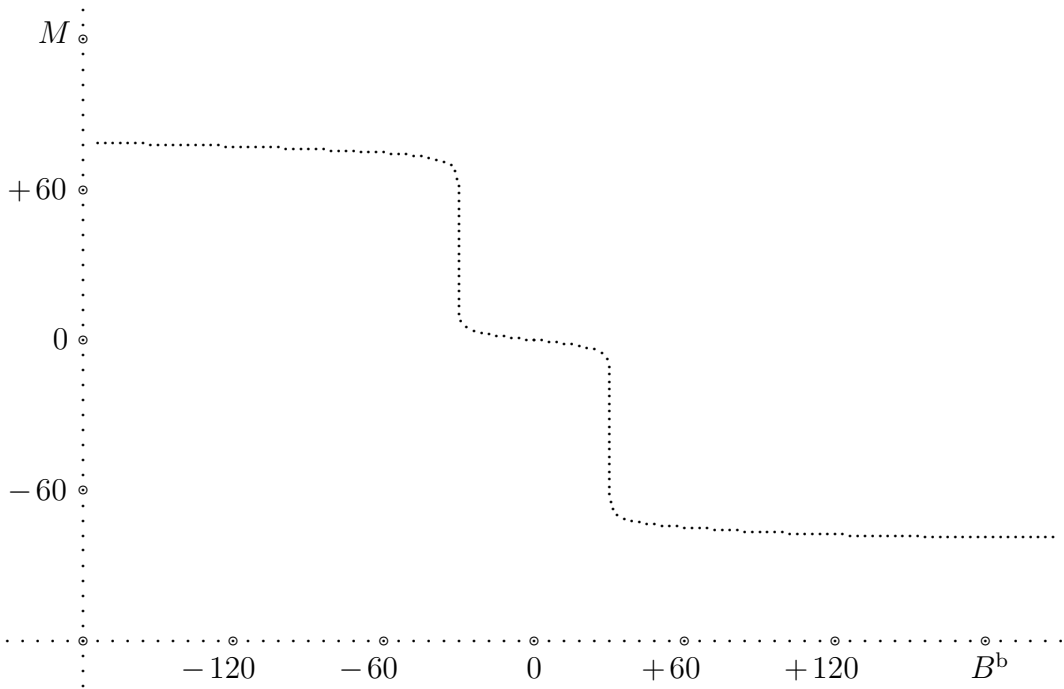
   Figure 6. $M$ $[\h Gauss]$ versus $B^{\rm b}$ $[\h Gauss]$
             for $(\pi n_e\hbar/e)\cdot\bar\alpha=30\h\h Gauss$
             and $T=40\h^{\circ}K$.
   \end{center}

\end{document}